\documentstyle[12pt]{article}
\begin{document}
\setlength{\unitlength}{1mm}
\title{{\hfill {\small Alberta-Thy 07-97} } \vspace*{2cm} \\
Mechanism of Generation of Black Hole Entropy
in Sakharov's Induced Gravity}
\author{\\
V.P. Frolov$^{1,2,3}$ and
D.V. Fursaev$^{1,4}$ \date{}}
\maketitle
\noindent  {
$^{1}${ \em
Theoretical Physics Institute, Department of Physics, 
\ University of Alberta, \\ Edmonton, Canada T6G 2J1}
\\ $^{2}${\em CIAR Cosmology and Gravity Program}
\\ $^{3}${\em P.N.Lebedev Physics Institute,  Leninskii Prospect 53,
Moscow 117924, Russia}
\\ $^{4}${\em Joint Institute for
Nuclear Research, Laboratory of Theoretical Physics, \\
141 980 Dubna, Russia}
\\
\\
e-mails: frolov, dfursaev@phys.ualberta.ca
}
\bigskip

\begin{abstract}
The mechanism of generation of the Bekenstein-Hawking  entropy $S^{BH}$
of a black hole in the Sakharov's induced gravity is proposed. It is
suggested that the "physical" degrees of freedom, which explain the
entropy $S^{BH}$, form only a finite subset of the standard 
Rindler-like modes defined outside the black hole horizon. The entropy
$S_R$ of the Rindler modes, or entanglement entropy, is always
ultraviolet divergent, while the entropy of the "physical" modes is
finite and it coincides in the induced gravity with $S^{BH}$. The two
entropies $S^{BH}$ and $S_R$ differ by a surface integral $Q$
interpreted as a Noether charge of non-minimally coupled scalar
constituents of the model. We demonstrate that energy $E$ and
Hamiltonian  $H$ of the fields localized in a part of space-time,
restricted by the Killing horizon $\Sigma$,  differ by the quantity
$T_H Q$, where $T_H$ is the temperature of a black hole.  The first law
of the black hole thermodynamics enables one to relate the probability
distribution of fluctuations of the black hole mass, caused by the
quantum fluctuations of the fields, to the probability distribution of
"physical" modes over energy $E$.  The latter turns out to be different
from the distribution of the Rindler modes.  We show that the
probability distribution of the  "physical"  degrees of  freedom has a
sharp peak at $E=0$  with the width proportional  to the Planck mass.
The logarithm of number of "physical" states at the peak  coincides
exactly with the black hole entropy $S^{BH}$. It enables us to argue
that the energy distribution of the "physical" modes and distribution
of the  black hole mass are equivalent in the induced gravity. Finally
it is shown that the Noether charge $Q$ is related to the entropy of
the low frequency modes propagating in the vicinity of the bifurcation
surface  $\Sigma$ of the horizon. We find in particular an explicit
representation of $Q$ in terms of an effective action of some 
two-dimensional quantum fields "living" on $\Sigma$.
\end{abstract}

\bigskip

{\it PACS number(s): 04.60.+n, 12.25.+e, 97.60.Lf, 11.10.Gh}

\baselineskip=.6cm

\newpage

\noindent
\section{Introduction}

Searching for statistical-mechanical explanation of the
Bekenstein-Hawking \cite{Beke:72}--\cite{Hawk:75} entropy $S^{BH}$ of
black holes attracted a lot of attention in the last years.  In
particular, one of the proposed ideas  was to relate $S^{BH}$ to
counting  of quantum excitations of a black hole 
\cite{Hooft:85}--\cite{BZ}. This suggestion, however, meets a
difficulty because the Bekenstein-Hawking entropy arises at tree-level
while the entropy of quantum excitations  is a one-loop  quantity. As
was first pointed out in  \cite{Jacobson}, this difficulty can be
resolved in the Sakharov's theory of induced  gravity
\cite{Sakh},\cite{Adler}.  According to the Sakharov's idea  general
relativity can be considered as a low energy effective theory where the
metric $g_{\mu\nu}$ becomes dynamical  variable as the result of
quantum effects in the system of heavy constituents propagating in the
external gravitational background.  Gravitons in this picture up to
some extend are  analogous to the phonon field describing collective 
excitations of a lattice in low-temperature limit of the theory. 

Surely, the Sakharov's approach does not provide us with a complete
understanding of gravity at Planckian scales and it cannot compete, for
instance, with superstring  models. Nevertheless, it has a number of
features, such as description of the graviton as a collective variable
and absence of the leading one-loop divergencies, which, in accordance
with our intuition, should be  the key properties of more profound
candidates to the role of quantum gravity theory.  One may hope, in
particular, that studying the Bekenstein-Hawking entropy in the
framework of the induced gravity would give us some hints how the
entropy can be explained in more realistic models and what is the
origin of its universality.

Recently we proposed a model \cite{FFZ3} which  illustrates explicitly
how the black hole entropy  $S^{BH}$ is generated in Sakharov's induced
gravity. Namely, it was shown that $S^{BH}$ is directly related to the 
statistical-mechanical entropy  $S_R$ of the thermally excited gas of
the heavy constituents (Rindler-like particles)  propagating in the
close vicinity of the black hole horizon
\begin{equation}\label{0.1}
S^{BH}=S_R-\bar{Q}~~~.
\end{equation}
Both fermions and bosons give positive and infinite contributions to
$S_R$,  so that this quantity is divergent. An additional term
$\bar{Q}$ in (\ref{0.1}) is  proportional to the fluctuations of the
non-minimally coupled scalar  fields $\hat{\phi}_s$ on the horizon
$\Sigma$ and is the average value  of the following operator
\begin{equation}\label{2.14}
\hat{Q}=2\pi\sum_{s}\xi_s \int_{\Sigma} 
\hat{\phi}_s^2 \sqrt{\gamma} d^2x~~~,
\end{equation}
where $\xi_s$ are the corresponding non-minimal couplings.  The
presence of such couplings is an important property of the model that
allows one to induce the gravitational action with the finite Newton
constant $G$.  The remarkable property of the model is that  for the
same values of the parameters of the constituents, that guarantee the
finiteness of  $G$, the divergences of $S_R$ are exactly cancelled by
the divergences of $\bar{Q}$. So in the induced gravity the
right-hand-side (r.h.s.) of (\ref{0.1}) is always finite and
reproduces  exactly the  Bekenstein-Hawking expression $S^{BH}={\cal
A}^H/(4G)$, where ${\cal A}^H$ is the surface area of the black hole.

Let us note that the operator $\hat{Q}$ has a clear interpretation as a
Noether charge. In Wald's approach  \cite{Wald:93} $\hat{Q}$ is that
part of the Noether charge which arises because of non-minimal
couplings of the scalar fields  $\hat{\phi}_s$ \cite{JKM}. This fact,
being virtually unimportant for classical black holes  which cannot
have scalar hair \cite{Mayo},  becomes crucial in the quantum theory
where fields have non-zero fluctuations $\langle\hat{\phi}_s^2
\rangle$. The Noether charge interpretation also gives us a hint how
the generalization of  formula (\ref{0.1}) for the black hole entropy 
might look in more realistic models.

Eq.(\ref{0.1}) shows that the entropy $S_R$ of the thermal bath outside
the black hole is much larger than the quantity $S^{BH}$ and thus it 
overcounts the degrees of freedom required to reproduce the
Bekenstein-Hawking entropy. Also, as has been pointed out by several
authors \cite{Solod:95}-\cite{Hotta},  on Ricci-flat geometries the 
entropy $S_R$ ignores non-minimal  couplings\footnote{By imposing on
the field near the horizon  special  boundary conditions depending on
$\xi$ one can make $S_R$ depending on $\xi$ as well.  Moreover recently
Solodukhin \cite{Solod:9612} suggested  such a scattering condition on
the horizon which enables one to reproduce the term $\bar{Q}$ in the
entropy (\ref{0.1}).  However the physical meaning of this scattering
condition is not clear.},  while massive fields contribution to
$S^{BH}$ depends on constants $\xi_s$.

\bigskip

In this paper we analyse the following two questions. First, what is
the mechanism which enables one to separate the "physical" modes
responsible  for the entropy $S^{BH}$ from other thermal excitations,
and, second,  what is the statistical-mechanical meaning of the
quantity $\bar{Q}$ in Eq.(\ref{0.1}).

\bigskip

We begin with the observation that  the Hamiltonian $H$ of a 
non-minimally coupled scalar field calculated for the black hole
exterior  differs from   energy $E$ calculated with the help of the 
stress-energy tensor $T_{\mu\nu}$. The difference $T_H Q$ is
proportional to the Noether charge $Q$, where $T_H$ is the Hawking
temperature of  the black hole.  The quantity $E$ defines the
difference between the mass of the black hole at the horizon and the
mass of the system measured at infinity. Thus for the fixed mass at
infinity  fluctuations of  quantum constituent fields result in the
quantum fluctuations of the black hole  mass.

On the other hand, the value of the Hamiltonian $H$ coincides with the
canonical energy.   For a fixed temperature Rindler-like modes of
constituents are thermally distributed with respect to the canonical
energy. Hence the Hamiltonian $H$ allows one to calculate the
statistical-mechanical entropy $S_R$ which enters Eq.(\ref{0.1}).
Formula (\ref{0.1}) indicates that  "physical" modes that are
responsible for the black hole entropy $S^{BH}$ form  a subset  of the
total set of the Rindler modes. We shall show that the difference
$\bar{Q}$ between $S_R$ and $S^{BH}$ is directly connected with the
difference between the energy $E$ and the canonical energy $H$ and is
defined by the fluctuations of the non-minimally coupled fields at the
horizon. It will be demonstrated that $\bar{Q}$ is completely
determined by zero-frequency ("soft") modes of scalar fields
propagating in the vicinity of the horizon. We shall also show  that
the statistical-mechanics of the soft modes  is equivalent to a
two-dimensional (2D)  quantum theory of  effective fields (fluctons) 
"living" on the bifurcation surface $\Sigma$.

We argue that the  leading temperature asymptotics of the canonical
ensembles of Rindler, "physical" and soft modes in the induced gravity
have a universal form because  they are determined only by the behavior
of the system near the horizon. It enables us to find for Ricci-flat
backgrounds the distribution of the "physical" degrees of freedom
explicitly. We show, in particular, that at the given temperature the
probability distribution of "physical" states has a sharp peak near the
average energy $E=0$ with the width determined by the masses of the
heaviest constituents of the  induced gravity models. The logarithm of
number  of physical states at the peak   is exactly the
Bekenstein-Hawking entropy.

\bigskip

Thus the proposed mechanism of the entropy generation in Sakharov's
induced gravity  implies that: i) the space of Rindler modes consists
of  subspaces of  "physical" and soft modes, and ii) the density number
of "physical" states at $E=0$ determines the degeneracy of the black
hole mass spectrum.  The obtained results confirm the consistency of
these suggestions.

\bigskip

The  paper is organized as follows. In Section~2 we formulate the model
of induced gravity and  recall some results of Ref.\cite{FFZ3} that are
necessary for our consideration.  In Section~3 we discuss a
non-minimally coupled scalar field defined on a part of black hole
background restricted by the  Killing horizon and find out the relation
between its energy and Hamiltonian. In Section~4 we show that the
fluctuations of a scalar field on  the bifurcation surface $\Sigma$ of
the  Killing horizons  is determined only by the contribution of the 
soft modes. We also obtain   the canonical thermal average of
$\langle\hat{\phi}^2 \rangle_\beta$ for the Rindler space. The relation
between  the degeneracy of the black hole mass spectrum and spectral
density of "physical" modes is discussed in Section~5. In Section~6 we
calculate the spectral density of Rindler states and discuss its
properties.  Spectral densities of the soft and "physical" modes are
found  in Section~7. These results are used to obtain the probability
distribution of the black hole mass and degeneracy of the black hole
spectrum. In Section~8 and Appendix we demonstrate that the statistical
mechanics of soft modes can be related to degrees of freedom of an
effective 2D theory. Namely, we show that the charge $\bar{Q}$ is
expressed in terms of a 2D effective action of some massive quantum
scalar fields "living" on the surface $\Sigma$. Section~9 contains
discussion of the results. 

We use sign conventions of the book \cite{MTW}, and thus use the
signature $(-,+,+,+)$ for a Lorentzian metric.

\section{Induced entropy of a black hole in Sakharov's induced gravity}
\setcounter{equation}0

We recall that the starting point of the induced gravity approach is
the equality
\begin{equation}\label{aa}
\exp(-W[g_{\mu\nu}])=\int {\cal D}\Phi_i 
\exp(-I[\Phi_i,g_{\mu\nu}])~~~,
\end{equation}
that expresses the effective action $W[g_{\mu\nu}]$ of the
gravitational low energy effective  theory in terms of a quantum
average of the constituent fields $\Phi_i$ propagating in a given
external gravitational background $g_{\mu\nu}$. The Sakharov's basic
assumption is that the gravity becomes dynamical only as the result of
quantum (one-loop) effects of the constituent fields.

A simple model convenient for the discussion of the problem of black
hole entropy in  induced gravity was suggested in 
\cite{FFZ3}\footnote{Another discussion of black hole entropy in
induced gravity can be found in \cite{Belgiorno:9612}.}. This model is
built of $N_s$ free scalar bosons $\phi_s$ with masses $m_s$ and of
$N_d$ free fermion fields $\psi_d$ with masses $m_d$.  The scalar
fields have non-minimal couplings  with constants $\xi_s$ and their
classical actions $I[\phi_s,g_{\mu\nu}]$ are similar to the scalar
action which will be considered in Section 3,  see Eq.(\ref{2.3}). The
fermion fields are the Dirac   spinors $\psi_d$ with the Dirac actions
$I[\psi_d,g_{\mu\nu}]$. Thus effective gravitational action
$W[g_{\mu\nu}]$ is defined by Eq. (\ref{1.3}) where the classical
action for the constituent fields $(\Phi_i=\{\phi_s,\psi_d\})$ is the
sum
\begin{equation}\label{i8}
I[\Phi_i,g_{\mu\nu}]=\sum_s I[\phi_s,g_{\mu\nu}]+
\sum_d I[\psi_d,g_{\mu\nu}]~~~.
\end{equation}

Consider now the following two functions 
\begin{equation}\label{c1}
p(z)=\sum_s m_s^{2z} -4\sum_d m_d^{2z}\, ,\hspace{.5cm}
q(z)=\sum_s m_s^{2z}(1-6\xi_s) +2\sum_d m_d^{2z}\, 
\end{equation}
constructed from the parameters of the constituents. Direct
calculations show that the induced cosmological  constant vanishes and
the induced gravitational coupling constant $G$ is finite if  the
following constraints are satisfied
\begin{equation}\label{c2}
p(0)=p(1)=p(2)=p'(2)=0\, , 
\end{equation}
\begin{equation}\label{c3}
q(0)=q(1)=0\, .
\end{equation}
In particular, the condition $p(0)=0$ requires that $N_s=4N_d$ and is
always satisfied in supersymmetric theories. The finite Newton constant
$G$ is the function of the parameters of the constituents
\begin{equation}\label{i10}
{1 \over G}={1\over 12\pi}q'(1)=
{1 \over 12\pi} \left(\sum_{s}(1-6\xi_s)~ m_s^2 \ln m_s^2+2\sum_{d}
m^2_d\ln m_d^2\right)~~~.
\end{equation}
For $N_d>1$ and $N_s>4$ the equations (\ref{c2}) for masses of the
constituents  are consistent, while equations (\ref{c3}) are linear
equations defining $\xi_s$.  Relation (\ref{i10}) shows that some of
the fields  (heavy constituents) have masses comparable to the Planck
mass. 

The heavy constituents are unobservable at low energies and their
effective action $W[g_{\mu\nu}]$ is reduced in the low energy regime to
the Einstein-Hilbert action
\begin{equation}\label{i11}
W[g_{\mu\nu}]=-{1\over 16\pi G}\left(\int_{\cal M} dV\, \, R 
+2\int_{\partial  {\cal M}}dv \, K\right)+\ldots \, \, .
\end{equation}
The dots in r.h.s. of (\ref{i11}) indicate higher curvature terms which
are suppressed  by the power factors of $m_i^{-2}$ when the curvature
is small. In order to make finite  the terms which are quadratic in
curvature one must consider a more general set of constituent fields
with additional constrains imposed on them. For our problem, since we
will be interested in Ricci-flat geometries, possible local ${\cal
R}^2$-terms  give a pure topological contribution to the action  which
is  irrelevant for our discussion. For this reason in what follows we
omit  such terms.

The variation of $W[g_{\mu\nu}]$ gives the Einstein equations
\begin{equation}\label{i12}
{\delta W\over {\delta g_{\mu\nu}}}\sim 
R^{\mu\nu}-{1\over 2}g^{\mu\nu} R=0\, .
\end{equation}
According to the Sakharov's equality (\ref{aa}) these equations in the
induced  gravity are equivalent to the relation
\begin{equation}\label{i13}
\left\langle \hat{T}^{\mu\nu} (x)\right\rangle =0\, ,
\end{equation}
where $ \hat{T}^{\mu\nu}(x)$ is the total stress-energy tensor of the
constituents.

The value of the Einstein-Hilbert (\ref{i11}) action calculated on the
Gibbons-Hawking instanton determines the classical free energy of the
black hole, and hence gives the Bekenstein-Hawking entropy. The
remarkable feature of the Sakharov's equality (\ref{aa}) is that it
allows one to rewrite identically the same classical free energy of the
black hole in terms of average over the  heavy constituents in the
Hartle-Hawking state  propagating on  the black hole background. By
using this representation it is possible to get an explicit expression
for the Bekenstein-Hawking entropy in terms of  constituents
\cite{FFZ3}
\begin{equation}\label{1.1}
S^{BH}=-\sum_{i}\mbox{Tr}~\hat{\rho}_i\ln\hat{\rho}_i-\sum_{s}
2\pi\xi_s\int_{\Sigma}\langle\hat{\phi}^2_s
\rangle \sqrt{\gamma} d^2x~~~.
\end{equation}
The thermal density  matrix $\hat{\rho}_i$ of the Rindler particles for
constituents  in the Hartle-Hawking state is
\begin{equation}\label{densitym}
\hat{\rho}_i={e^{-\beta_H{\hat H}_i}
\over \mbox{Tr}~e^{-\beta_H{\hat H}_i}}~~~.
\end{equation}
Here ${\hat H}_i$ are Hamiltonians of the fields and $\beta_H$ is the
inverse Hawking temperature. Equation (\ref{1.1}) shows that $S^{BH}$
is related to the statistical-mechanical entropy
$S_R=-\sum_i\mbox{Tr}~\hat{\rho}_i\ln\hat{\rho}_i$ of the heavy
constituents computed on the given black hole  background\footnote{The
entropy $S_R$ can be also interpreted as an entanglement entropy, see
\cite{Sorkin},\cite{Srednicki},\cite{CW}.}, and also depends on the
average of the square of the scalar field operators on the black hole
horizon $\Sigma$.

Note that each term  $-\mbox{Tr}~\hat{\rho}_i\ln\hat{\rho}_i$ in
(\ref{1.1}) is positive and divergent. The last term in the r.h.s. of
(\ref{1.1}) appears because of non-minimal couplings. Presence of such
couplings  in the scalar sector of the model is imperative in order  to
provide the ultraviolet finiteness of the Newton constant in  the
low-energy gravitational action.  What is remarkable, the terms with
non-minimal couplings  exactly cancel all the divergencies in $S_R$ so
that r.h.s of (\ref{1.1}) correctly reproduces the Bekenstein-Hawking
entropy.

Strictly speaking the Bekenstein-Hawking entropy is only the leading
part of the r.h.s. of (\ref{1.1}). Since the curvature ${\cal R}$ of
the spacetime does not vanish relation (\ref{1.1}) also contains the
corrections of the order $m^{-2}_i{\cal R}$. These terms are directly
related to the higher order in curvature corrections to the
Einstein-Hilbert action (\ref{i11}). The masses of the constituents are
very high and their modes are thermally excited only in the narrow
region in the vicinity of the horizon. For the description of these
modes we shall use the Rindler approximation and omit the curvature
dependent corrections to $-\mbox{Tr}~\hat{\rho}_i\ln\hat{\rho}_i$ and
$\langle \hat{\phi}_s^2\rangle$ that are of the same order
$m^{-2}_i{\cal R}$ as the terms omitted in  (\ref{i11}).

After these remarks let us discuss the concrete mechanism of
cancellation of divergences in (\ref{1.1}). The partition function for
a massive field with the mass $m_i$ can be calculated explicitly in the
limit when the curvature radius of the space-time is much larger then
the Compton wave length of the field
\begin{equation}\label{1.3}
\mbox{Tr}~e^{-\beta\hat{H}_i}\simeq
\exp(-\mu_i\beta-U_i(\beta))
\end{equation}
(see for the details Ref.\cite{FFZ3}). Here $\beta^{-1}$ is the
temperature of the system measured at infinity, $\hat{H}_i$ is the
Hamilton  operator of the field in question,  $\mu_i$ is a parameter
associated to the vacuum energy and 
\begin{equation}\label{1.4}
U_i(\beta)=-g(m_i^2)~{\pi \over 6} 
{\beta_H \over \beta} {\cal A}^H~~~.
\end{equation}
Here ${\cal A}^H=\int_{\Sigma} \sqrt{\gamma} d^2x$ is the area of the 
horizon.  The function $g(m_i^2)$ depends on the mass $m_i$ of the
field and it is  given by the integral
\begin{equation}\label{1.5}
g(m_i^2)=
n_i\int_0^{\infty} {ds \over (4\pi s)^{D/2}}e^{-m_i^2s}~~~,
\end{equation}
where $D$ is the dimensionality of the space time and the factor $n_i$
is equal to 1 or 2 for scalars and  (4D Dirac) fermions, respectively.
This integral is divergent and it has to be regularized  by using, for
instance, the Pauli-Villars \cite{DLM} or the dimensional \cite{FFZ3} 
regularizations. The function $g(m_i^2)$ is important because it
determines the statistical-mechanical entropy 
$-\mbox{Tr}~\hat{\rho}_i\ln\hat{\rho}_i$ of the  given field evaluated
at $\beta=\beta_H$
\begin{equation}\label{1.6}
-\mbox{Tr}~\hat{\rho}_i\ln\hat{\rho}_i
=\left(1-\beta {\partial \over \partial \beta}\right)
\ln \mbox{Tr}~e^{-\beta\hat{H}_i}=
g(m_i^2){\pi \over 3}{\cal A}^H~~~.
\end{equation}
On the other hand, the same function $g(m_i^2)$ determines the average
value of the scalar field $\langle\hat{\phi}_s^2\rangle$ on  the
horizon 
\begin{equation}\label{1.7}
\int_{\Sigma}\langle\hat{\phi}_s^2\rangle\sqrt{\gamma}d^2x
=
g(m_s^2){\cal A}^H~~~.
\end{equation}
Now, if  entropies $-\mbox{Tr}~\hat{\rho}_i\ln\hat{\rho}_i$ and
averages (\ref{1.7}) are regularized according to the same scheme with
the equal regularization parameters, the substitution of
Eqs.(\ref{1.6}) and (\ref{1.7}) into Eq.(\ref{1.1}) gives the
Bekenstein-Hawking entropy in the induced gravity 
\begin{equation}\label{xxx}
S^{BH}= {1 \over 4G}
{\cal A}^H~~~,
\end{equation}
\begin{equation}\label{yyy}
 {1 \over G}=
{4\pi \over 3} \left(\sum_s g(m^2_s)(1-6\xi_s)
+\sum_d g(m^2_d)\right)~~~. 
\end{equation}
The Newton constant defined by (\ref{yyy}) is ultraviolet finite,
provided  the constraints (\ref{c2}) and (\ref{c3}) are satisfied, and
after regularization is removed its expression is given by
Eq.(\ref{i10}).  Formulas (\ref{xxx}), (\ref{yyy})  explicitly
demonstrate the crucial role of the non-minimal coupling. It shows, in
particular, that the total  contribution of quantum fields to the black
hole entropy can be finite only if  $\xi_s>0$ for some constituents.

\section{Energy and Hamiltonian}
\setcounter{equation}0

We discuss now how the non-minimal coupling of scalar  constituents
manifests itself in the black  hole thermodynamics. For simplicity in
what follows we consider a static black hole.  The consideration can be
easily extended to the stationary case as well.

Let us recall that the complete background spacetime of an eternal
black hole contains  two Rindler-like wedges bounded by the Killing
horizons. The Killing horizons intersect at the two dimensional 
bifurcation  surface $\Sigma$. We shall use a foliation of the
hypersurfaces $t$=const orthogonal to the Killing vector $\zeta^\mu$
that intersect each other at $\Sigma$. 

Quantum fields propagate  on the complete  space-time manifold, 
however in our statis-\\tical-mechanical calculations we  restrict
ourselves by considering only a part of the system located in one of
the wedges. It is instructive first to discuss how this procedure
manifests in the classical theory. We focus on a classical scalar field
$\phi$ with a non-minimal coupling with the scalar curvature $R$
described by the action
\begin{equation}\label{2.3}
I[\phi]=-\frac 12\int(\phi^{,\mu}\phi_{,\mu}+m^2\phi^2
+\xi R\phi^2)\sqrt{-g}~d^4x~~~.
\end{equation}
The field obeys the  equation
\begin{equation}\label{2.1}
{\,\lower0.9pt\vbox{\hrule \hbox{\vrule height 0.2 cm 
\hskip 0.2 cm \vrule height 0.2 cm}\hrule}\,}
\phi-(m^2+\xi R)\phi=0~~~,
\end{equation}
where
${\,\lower0.9pt\vbox{\hrule \hbox{\vrule height 0.2 cm 
\hskip 0.2 cm \vrule height 0.2 cm}\hrule}\,}$ is the 
D'Alambert operator. The  stress-energy tensor resulting from the
variation of the action (\ref{2.3}) with respect to the metric is
\begin{equation}\label{2.2}
T_{\mu\nu}=\phi_{,\mu}\phi_{,\nu}-\frac 12 g_{\mu\nu}
\left(\phi_{,\rho}\phi^{,\rho}+m^2\phi^2\right)
+\xi\left[(R_{\mu\nu}-\frac 12 g_{\mu\nu} R)\phi^2
+g_{\mu\nu} (\phi^2)^{,\rho}_{~;\rho}-
(\phi^2)_{;\mu\nu}\right] \, .
\end{equation}

Denote by ${\cal B}$ a space-like hypersurface  orthogonal to the 
Killing vector $\zeta^\mu$. The  energy $E$ of the system is defined in
terms of the stress-energy  tensor (\ref{2.2})
\begin{equation}\label{2.4}
E=\int_{{\cal B}} \ T_{\mu\nu}\zeta^{\mu}d\sigma^{\nu}=-
\int_{{\cal B}}T^0_0\sqrt{-g}~d^3x~~~,
\end{equation}
where $d\sigma^{\nu}$ is the future directed vector of the volume 
element on ${\cal B}$. In the general case $E$ differs from  the {\it
canonical} energy $H$  that coincides with the Hamiltonian. The latter
is expressed in terms of the Hamiltonian density  ${\cal H}$
\begin{equation}\label{2.5}
H=\int_{{\cal B}} {\cal H}\sqrt{-g}~d^3x~~~,
\end{equation}
where
\begin{equation}\label{2.6}
{\cal H}=\frac 12\left(-g^{00}\phi_{,0}^2 
+g^{ij}\phi_{,i}\phi_{,j}+(m^2+\xi R)\phi^2\right)~~~.
\end{equation}
To compare $E$ and $H$ we note that in a static space-time
\begin{equation}\label{2.7}
-T^0_0={\cal H}
-\xi\left(R^0_0\phi^2+g^{ij}(\phi^2)_{;ij}
\right)~~~.
\end{equation}
The last term in r.h.s. of this equation can be rewritten as 
$$
(\phi^2)_{;ij}=
{\,\lower0.9pt\vbox{\hrule \hbox{\vrule height 0.2 cm 
\hskip 0.2 cm \vrule height 0.2 cm}\hrule}\,}\phi^2-
g^{00}(\phi^2)_{;00}=g^{00}\left((\phi^2)_{,0,0}
-(\phi^2)_{;00}\right)+
{1 \over \sqrt{-g}}\partial_i\left(\sqrt{-g}g^{ij}\partial_j
\phi^2\right)=
$$
\begin{equation}\label{2.8}
=
{1 \over \sqrt{-g}}\partial_i\left(\sqrt{-g}g^{ij}
((\phi^2)_{,j}-\phi^2w_j)\right)+\nabla_{\mu}w^{\mu} 
\phi^2
~~~.
\end{equation} 
Here $w^{\mu}=\frac 12 \nabla^{\mu}\ln |g_{00}|$ is a time-independent 
acceleration of the Killing observer.  It can be shown that
$\nabla_{\mu}w^{\mu}=-R^0_0$, so that for static space-times  relation
(\ref{2.7}) takes the form
\begin{equation}\label{2.7b}
-T^0_0={\cal H}-\xi{1 \over \sqrt{-g}}\partial_i\left(\sqrt{-g}g^{ij}
((\phi^2)_{,j}-\phi^2w_j)\right)~~~.
\end{equation}
Then substitution of (\ref{2.7b}) into (\ref{2.4}) gives the required
relation between the energy $E$ and  the canonical energy $H$
\begin{equation}\label{2.9}
E=H-\xi\int_{\partial {\cal B}}ds^k~|g_{00}|^{1/2}
((\phi^2)_{,k}-\phi^2w_k)~~~.
\end{equation}
Here  $ds^k$  is a three dimensional vector in ${\cal B}$ normal  to
the boundary $\partial {\cal B}$ and directed outward with respect to
${\cal B}$. Thus two energies differ by a surface term given on the
boundary $\partial {\cal B}$ of the hypersurface ${\cal B}$.

Obviously, when one considers a complete Cauchy surface  the boundary
term in (\ref{2.9}) contains only a contribution from the spatial
infinity, or from the external spatial boundaries if they are present.
For a field falling off  at infinity or obeying suitable conditions at
the boundary one can get rid of the boundary term and make $E$ and $H$
be equal.

However, the situation is qualitatively different when we consider the
theory only in one of the wedges. Then the integration region in $E$ is
restricted by the bifurcation surface $\Sigma$ of the Killing horizon,
where the field $\phi$ can take arbitrary finite values. By assuming
that contribution from the spatial infinity or external boundary is
absent  one can write for the field regular at the horizon the
following relation
\begin{equation}\label{2.10}
E=H-\beta_H^{-1} Q~~~,
\end{equation}
were $\beta_H^{-1}$ is the  Hawking temperature  determined by  the
surface gravity $\kappa$ of the black hole as
$\beta_H^{-1}=\kappa/(2\pi)$ and
\begin{equation}\label{2.10a}
Q=2\pi \xi \int_{\Sigma}\phi^2\sqrt{\gamma}
d^2x~~~.
\end{equation}
We see therefore that the energy $E$  computed for a domain restricted
by the horizon differs from the canonical energy $H$ by the quantity 
proportional to $Q$.

It is reasonable to ask what is the relevance of  quantities $E$ and
$H$ from the point of view of  black hole thermodynamics. To this aim
one can consider the field $\phi$ on a black hole background and
calculate its contribution to the entropy and energy of a black hole. 
A simple way to do this is to make use of Euclidean  formulation of the
theory on the black hole instanton with the arbitrary period $\beta$ of
the Euclidean time. The results obtained in such an off-shell approach
coincide with the results of other methods 
\cite{Nelson}-\cite{WaldIyer:95}. If $\beta\ne\beta_H$ the background
has a conical singularity and one can write the Euclidean action in the
form\footnote{ As earlier we omit the terms connected with the external
boundary that might be present. They are not important for our
consideration and, as was explained, they can be avoided by choosing
the appropriate boundary conditions.}
\begin{equation}\label{2.11}
I_E[\phi,g_{\mu\nu},\beta]=\frac 12\int_{{\cal M}_\beta}
\left(\phi^{,\mu}\phi_{,\mu}+m^2\phi^2+\xi R\phi^2
\right)\sqrt{-g}~d^4x+2\pi\xi\left(1-{\beta \over \beta_H}
\right)\int_{\Sigma}\phi^2 \sqrt{\gamma}d^2x~~~.
\end{equation}
Since for a static field configuration the bulk part of  Euclidean
action (\ref{2.11}) is proportional to  Hamiltonian  (\ref{2.6}) one
can rewrite Eq.(\ref{2.11}) as
\begin{equation}\label{2.11b}
I_E[\phi,g_{\mu\nu},\beta]=
\beta H +2\pi\xi\left(1-{\beta \over \beta_H}
\right)\int_{\Sigma}\phi^2\sqrt{\gamma}d^2x~~~.
\end{equation}

To derive the contributions $\Delta E$ and $\Delta S$  of the scalar
field to the mass and entropy of the black hole one just identifies
$\beta^{-1}$ with the temperature and functional
$\beta^{-1}I_{E}(\beta)$ with the free energy. If the scalar field does
not vanish at the horizon one has
\begin{equation}\label{2.12}
\Delta E={\partial \over \partial \beta}I_E[\phi,g_{\mu\nu},\beta]
|_{\beta=\beta_H}
=H- \beta_H^{-1} Q~~~,
\end{equation}
\begin{equation}\label{2.13}
\Delta S=\left(\beta{\partial \over \partial \beta}
-1\right)I_E[\phi,g_{\mu\nu},\beta]|_{\beta=\beta_H}
=-Q~~~,
\end{equation}
where $Q$ is defined by Eq.(\ref{2.10a}). Thus the  energy $\Delta E$
appears as a part of total energy of the system. As one can see from
(\ref{2.12}) and (\ref{2.13}) the  quantity $Q$ contributes  both to
the mass and entropy of the black hole. Yet  in the classical theory
such contributions  are absent when the  field and metric obey the
classical equations. This conclusion follows from the recent analysis
by Mayo and Bekenstein \cite{Mayo} who  showed that for any value of
the non-minimal coupling $\xi$ a stationary black hole   has no massive
scalar hair.  

The situation is quite different for a quantum field. Due to the
presence of vacuum zero-point  fluctuations the average $\langle
\hat{\phi}^2 \rangle$ does not vanish. That is why quantum 
fluctuations of scalar fields on $\Sigma$  manifest themselves in the
black hole thermodynamics. Moreover, the contribution of a quantum 
scalar field to black hole entropy (\ref{1.1}) because of non-minimal
coupling directly follows from  (\ref{2.13}) if one  replaces the
classical quantity $Q$ by its quantum version $\bar{Q}\equiv\langle
\hat{Q} \rangle$ and takes the sum over all  non-minimally coupled 
constituents.

\section{Soft modes} \setcounter{equation}0

Our aim now is to investigate the properties of the quantity $Q$.
Because $Q$ is defined strictly on $\Sigma$ it is sufficient to
consider only the behavior of scalar fields in the domain close to the
horizon surface where the black hole metric can be  approximated by the
Rindler metric
\begin{equation}\label{3.1}
ds^2=-\kappa^2\rho^2 dt^2+d\rho^2+(dz^1)^2+(dz^2)^2~~~,
\end{equation}
where $\kappa=2\pi/\beta_H$. Our aim is to demonstrate that the value
of $Q$ is determined only by a contribution of Rindler modes with 
negligibly small  frequencies. 

We begin with the analysis of $Q$ in classical theory. A normalized
solution of the classical Klein-Gordon equation in the Rindler space is
\cite{Takagi}
\begin{equation}\label{3.2}
U_{\omega,k}(x)={1 \over 2\pi} u_{\omega,k}(t,\rho)e^{-ik_jz^j} ~~~,
\end{equation}
\begin{equation}\label{3.2a}
u_{\omega,k}(t,\rho)
={1 \over 2\pi ^2}(\sinh \pi \omega)^{1/2}~
K_{i\omega}(\mu\rho)e^{-i\kappa \omega t}~~~,
\end{equation}
where  $\omega$ is the dimensionless frequency,
$\mu=(m^2+k_j^2)^{1/2}$, $j=1,2$,  and  $K_{i\omega}(x)$ is the
modified (hermitean) Bessel function which vanish at
$\rho\rightarrow\infty$.  An interesting observation concerning these
modes is that  only modes with negligibly small frequencies $\omega$
contribute to the value of the  field $\phi$ on the Rindler horizon. We
call such solutions {\it soft} modes. Their behavior near the horizon
follows from the asymptotic of the modified Bessel functions at small
values of $\mu\rho$
\begin{equation}\label{asympt}
K_{i\omega}(\mu\rho)\simeq {i\pi \over 2\sinh \pi \omega}
\left[{1 \over \Gamma(i\omega+1)}
\left({\mu\rho \over 2}\right)^{i\omega}
-{1 \over \Gamma(-i\omega+1)}
\left({\mu\rho \over 2}\right)^{-i\omega}
\right]~~~.
\end{equation}
By using the formula
\begin{equation}\label{limit}
\lim_{a\rightarrow 0}{\sin(x\ln a) \over x}= 
-{\pi \over 2} \delta(x)~~~,
\end{equation}
where the delta function is normalized on the half axis, one can define
the limiting value of $K_{i\omega}(\mu\rho)$ as  the distribution
\begin{equation}\label{3.3}
\lim_{\rho\rightarrow 0}K_{i\omega}(\mu\rho)=
{\pi \over 2}\delta(\omega)~~~.
\end{equation}
The consequence of (\ref{3.2}) and (\ref{3.3}) is that in the presence
of the non-minimal coupling classical energy and Hamiltonian are
affected by  the soft  modes in the different way. Consider, for
example,  a wave packet $\phi_{\triangle \omega}(t,\rho,z)$ which is
constructed of soft modes with frequencies $\omega$ in the range
$(0,\triangle \omega)$.  Let function $\phi_{\triangle
\omega}(t,\rho,z)$ be a solution of the Klein-Gordon equation of the
form
\begin{equation}\label{3.4}
\phi_{\triangle \omega}
(t,\rho,z)= {1  \over 2\pi} 
\int_{0}^{\infty}d\omega\int d^2k~ u_{\omega,k}(t,\rho) e^{-ik_jz^j}
\tilde{\phi}_{\triangle \omega}(\omega)\tilde{\varphi}(k)~~~.
\end{equation}
Here $\tilde{\phi}_{\triangle \omega}(\omega)$ and $\tilde{\varphi}(k)$
are some functions of $\omega$ and $k_j$, and
\begin{equation}\label{3.7}
\tilde{\phi}_{\triangle \omega}(\omega)=0~~~,
~~~\mbox{if}~~\omega>\triangle \omega~~~.
\end{equation}
To have a non-zero value $\phi_{\triangle \omega}(t,\rho,z)$ 
at $\rho\rightarrow 0$ we assume that
\begin{equation}\label{3.6}
\tilde{\phi}_{\triangle \omega}(\omega)
\simeq{2 \over \sqrt{\pi \omega}}~~~
,~~~\mbox{when}~~\omega\rightarrow 0~~~.
\end{equation}
Then 
\begin{equation}\label{3.9}
\lim_{\rho\rightarrow 0}\phi_{\triangle \omega}(t,\rho,z)=
\varphi(z)
~~~,
\end{equation}
where $\varphi(z)$ is defined as 
$$
\varphi(z)= {1 \over 2\pi}
\int d^2k~e^{-ik_iz^i} \tilde{\varphi}(k)~~~.
$$
The canonical energy of this wave packet, given by the integral
\begin{equation}\label{3.8}
H[\phi_{\triangle \omega}]=\int_{0}^{\infty}d\omega~\omega~ 
|\tilde{\phi}_{\triangle \omega}(\omega)|^2 
\int d^2 z |\varphi(z)|^2~~~,
\end{equation}
can be made arbitrary small as  $\triangle \omega\rightarrow 0$. On the
other hand, the value of $\phi_{\triangle \omega}$ on the horizon does
not vanish so that the energy $E$ of the wave packet
\begin{equation}\label{3.10}
E[\phi_{\triangle \omega}]=-2\pi\xi \beta_H^{-1}\int d^2 z 
|\varphi(z)|^2~~~
\end{equation}
remains non-zero.

\bigskip

We demonstrate now how the soft modes generate the Noether charge
$\bar{Q}$ in quantum theory. Consider the quantum scalar field and
calculate its correlator on the bifurcation surface $\Sigma$
($\rho=0$). The correlator for the canonical ensemble of the Rindler
particles at the temperature $\beta^{-1}$ is defined as
\begin{equation}\label{3.11}
G_\beta(x,x')=\langle \hat{\phi}(x)\hat{\phi}(x')\rangle_\beta
=\mbox{Tr}\left[\hat{\rho}(\beta)
\hat{\phi}(x)\hat{\phi}(x')\right]~~~,
\end{equation}
where $\hat{\rho}(\beta)$ is the density matrix  (\ref{densitym})
(where $\beta_H$ is replaced by $\beta$). Expression (\ref{3.11}) can
be rewritten as
\begin{equation}\label{3.12}
\langle \hat{\phi}(x)\hat{\phi}(x')\rangle_\beta=
\int_{0}^{\infty}d\omega\int d^2k\left[
n_\omega(\beta)~ U^{*}_{\omega,k}(x)U_{\omega,k}(x')+
(n_\omega(\beta)+1)U_{\omega,k}(x)U^{*}_{\omega,k}
(x')\right]~,
\end{equation}
where $n_\omega(\beta)$ is the density of particles with the energy
$\omega$ 
\begin{equation}\label{3.13}
n_\omega(\beta)=\left(e^{\kappa \beta 
\omega}-1\right)^{-1}~~~.
\end{equation}
Eq.(\ref{3.12}) follows from the decomposition of field operators in
the Rindler basis
\begin{equation}\label{3.14}
\hat{\phi}(x)=\int_{0}^{\infty}d\omega\int d^2k\left[
U^{*}_{\omega,k}(x)
\hat{b}^{+}(\omega,k)
+U_{\omega,k}(x)\hat{b}(\omega,k)\right]~~~,
\end{equation}
and formula
\begin{equation}\label{3.15}
\langle \hat{b}^{+}(\omega,k)\hat{b}(\omega',k')\rangle_\beta=
\delta(\omega-\omega')\delta^{(2)}(k-k')
n_\omega(\beta)~~~,
\end{equation}
where $\hat{b}^{+}(\omega,k)$ and $\hat{b}(\omega,k)$ are creation and
annihilation operators of the Rindler particles. The restriction of the
correlator $G_\beta(x,x')$ on the bifurcation surface $\Sigma$ is
obtained in the limit when coordinates $\rho$ and $\rho'$ of its both
points $x$ and $x'$ tend to zero. According to Eq.(\ref{3.3}),  the
main contribution  to the correlator in this limit is given by the
modes with negligibly small frequencies $\omega$. The density number of
such modes  is singular and is approximated by the expression
$$
n_\omega(\beta)\simeq {1 \over \kappa \beta \omega}~~~.
$$
So one finds for small $\rho$ and $\rho'$
\begin{equation}\label{3.16}
G_\beta(x,x')\simeq {1 \over 2\pi^4\kappa \beta}
\int d^2k e^{ik(z-z')}\int_{0}^{\infty}d\omega
{\sinh \pi \omega \over \omega}
K_{i\omega}(\mu\rho)K_{i\omega}(\mu\rho')~~~.
\end{equation}
The integration over $\omega$ can be done by making use  of asymptotic
(\ref{asympt}) and property (\ref{limit}) and the result reads
\begin{equation}\label{3.17}
\int_{0}^{\infty}d\omega
{\sinh \pi \omega \over \omega}
K_{i\omega}(\mu\rho)K_{i\omega}(\mu\rho')
\simeq -{\pi ^2 \over 4}\ln (\mu^2\epsilon^2)~~~.
\end{equation}
Here $\epsilon$ is a constant with the dimensionality of a length,
which is introduced to keep the  expression in the logarithm
dimensionless. In derivation of (\ref{3.17}) we omitted terms  which do
not depend on $\mu$.  These terms give a contribution to
$G_\beta(x,x')$  on $\Sigma$ proportional to $\delta^{(2)}(z-z')$ and
vanishing when $z\neq z'$. Denote by ${\cal G}_\beta(z,z')$ the
limiting value of $G_\beta(x,x')$ on  the bifurcation surface, where
$z$ and $z'$ are the coordinates of the points $x$ and $x'$ on
$\Sigma$. Because $\mu^2=m^2+k_j^2$ one finds from (\ref{3.16}) the
following expression 
\begin{equation}\label{3.18}
{\cal G}_\beta(z,z') =-{1 \over 2(2\pi)^2\kappa \beta}
\int d^2k~ e^{ik_j(z-z')^j}
\ln[(m^2+k_j^2)\epsilon^2] ~~~.
\end{equation}
When the arguments $z$ and $z'$ coincide  the integral over $k$ in
expression (\ref{3.18}) has to be regularized. This is a standard
problem when one is dealing with the  coincidence limit of Green
functions. By assuming that such a regularization is carried out we
find that the quantity  ${\cal G}_\beta(z,z)$ determines a non-zero
canonical average of the charge $\hat{Q}$
\begin{equation}\label{3.201}
\langle\hat{Q}\rangle_\beta =2\pi\xi \int_{\Sigma}
\langle\hat{\phi}^2(z)\rangle_\beta ~d^2z=
2\pi\xi \int_{\Sigma}
{\cal G}_\beta(z,z)d^2z~~~. 
\end{equation}

The results of our analysis can be summarized in the  following way. In
classical theory the quantity $Q$ is determined by the Rindler modes
with zero frequencies.   This property also holds in the quantum
theory  where the average $\langle\hat{Q}\rangle_\beta$ is not zero
because the density $n_\omega(\beta)$ of  the low-frequency modes is
singular\footnote{Note that this is a specific property of the boson
fields and this is not true for the fields with Fermi statistics.}. In
fact one can conclude that only soft modes with  $\omega \ll 1$ are 
responsible for the non-zero value of correlator (\ref{3.18}).

\section{Spectral density of states of a black hole in induced gravity}
\setcounter{equation}0

We return now to the problem of the statistical-mechanical origin of
the black hole entropy in the induced gravity. Since the entropy of a
black hole of mass $M$ is $S^{BH}=4\pi M^2/G$, one might expect that
the density of states of such a black hole is
\cite{Hawk:76},\cite{York:86}
\begin{equation}\label{a1}
\nu_{BH}(M)\Delta M\sim \exp({4\pi M^2\over G})\Delta M \, .
\end{equation}
The  statistical-mechanical foundation of black hole thermodynamics
implies the explanation of this degeneracy. Let us discuss how this
problem can be solved in the induced gravity.  We shall see that soft
modes introduced in the previous Section play an important role in this
discussion.

In the "constituent representation"  of the gravitational action we are
dealing with the  ultraheavy particles propagating in the given
external background.  The average energy of these particles is
\begin{equation}\label{a2}
\bar{E}=\left\langle E \right\rangle=\int \left\langle T_{\mu\nu}
 \right\rangle\zeta^{\mu}d\sigma^{\nu}=0\, .
\end{equation}
Hence the mass $M_{BH}$ of the black hole measured at the horizon and
the mass  $M_{\infty}$ measured at infinity are the same.  This
equality is the result of averaging over the states of the constituent
fields. A particular mode of a quantum field  contributes to the
energy, and hence to the mass. This gives rise to the difference
\begin{equation}\label{a3}
M_{\infty}-M_{BH}=\Delta M=\Delta E=\int  T_{\mu\nu} 
 \zeta^{\mu}d\sigma^{\nu}\, 
\end{equation}
determined by the differential mass formula \cite{BCH}. If one fixes
the  mass of the system measured at infinity the mass of  the black
hole is not fixed but fluctuates near its average value. The origin of
these fluctuations  (black hole mass "zitterbewegung") are the
fluctuations of the quantum constituent fields in  vicinity of the
black hole.  Moreover in the induced gravity the  degeneracy (\ref{a1})
of the black hole can be related  to the number $\nu(E)$ of physically
different states of constituents propagating in the black hole exterior
and having the total energy  $E$ in the interval $(0,\Delta M)$.
Namely, we have
\begin{equation}\label{a4}
\nu_{BH}(M)=\nu(E=0)\, .
\end{equation}
The problem of counting the states of constituents is technically much
simpler.  We suggest that "physical" states together with soft mode
states form a complete set that can be identified with the space of
Rindler modes. For this reason in order to obtain $\nu(E)$ it is
sufficient to find the corresponding spectral densities of states for
the Rindler and soft modes. Strictly speaking to calculate the density
of states one needs to know the properties of the system at the
temperature different from  the Hawking value, i.e. one  must consider 
the so called "off-shell" configurations.  However the "off-shellness"
can be arbitrary small, and we shall see that only the properties of
the states near the Hartle-Hawking equilibrium are really important.

\section{Spectral density of Rindler states} \setcounter{equation}0

We begin with analysis of the properties of the thermal canonical
ensembles of Rindler  particles\footnote{Since we are dealing with  a
black hole the term "Boulware"  might be more appropriate. We use a
term "Rindler"  both following tradition and in  order to stress the
nature of  our approximation.}.  Let $Z_{R}(\beta)$ be the
statistical-mechanical partition function of the complete set of scalar
and spinor constituents of the induced gravity model propagating in the
Rindler-like wedge
\begin{equation}\label{6.2}
Z_{R}(\beta)=\mbox{Tr}~e^{-\beta 
\hat{H}}~=~\prod_{i}\mbox{Tr}_i~e^{-\beta 
\hat{H}_i}~~~.
\end{equation}
Here $\hat{H}=\sum\hat{H}_i$ is the total Hamiltonian, and $\hat{H}_i$
are the Hamilton operators for the each particular constituent. For the
Hartle-Hawking vacuum  $\beta=\beta_H$.  We shall use subscript $R$ to
refer to quantities that are obtained by means of  $Z_R(\beta)$.

The  energy $E_{R}$ and entropy $S_{R}$ of the Rindler modes are
defined as
\begin{equation}\label{6.3}
E_R=-{\partial \over \partial \beta}
\ln Z_R(\beta)|_{\beta=\beta_H}~~~,
\end{equation}
\begin{equation}\label{6.4}
S_R=-\left(\beta
{\partial \over \partial \beta}-1\right)
\ln Z_R(\beta)|_{\beta=\beta_H}~~~.
\end{equation}
Another representation for these quantities is
\begin{equation}\label{6.5}
E_R=\left\langle \hat{H} \right\rangle=
\sum_i\left\langle \hat{H}_i \right\rangle\equiv
\sum_i\mbox{Tr}_i~(\hat{\rho}_i \hat{H}_i)~~~,
\end{equation}
\begin{equation}\label{6.6}
S_R=-\sum_i\mbox{Tr}_i~\hat{\rho}_i\ln
\hat{\rho}_i~~~.
\end{equation}
In (\ref{6.5}) we made use of the thermalization theorem \cite{Takagi}
according to that the statistical-mechanical average at $\beta=\beta_H$
of the operators localized in the wedge is equivalent to the
quantum-mechanical average in the Hartle-Hawking vacuum.

The Rindler partition function $Z_R(\beta)$ can be found explicitly
\cite{FFZ3} and it is given by  Eqs.(\ref{1.3}) and (\ref{1.4}).  It
can be presented as
\begin{equation}\label{6.91}
Z_R(\beta)=\exp(-\beta \mu_R+\beta^{-1}\lambda_R)~~~.
\end{equation}
By making use of Eqs.(\ref{1.3}) and (\ref{1.4}) one finds that
\begin{equation}\label{6.92}
\lambda_R={\beta_H \over 2} S_R={1\over 3}g (4\pi M)^3 ~~~,
\end{equation}
\begin{equation}\label{6.93}
E_R-\mu_R={1 \over 2\beta_H} S_R={\pi\over 3}g M ~~~,
\end{equation}
where $g=\sum_i g(m^2_i)$.

The spectral density $\nu_R(E)$ of the operator $\hat{H}_{R}$ is
defined in the standard way
\begin{equation}\label{10.2}
\nu_R(E)=\mbox{Tr}_R~\delta(\hat{H}-E)=
{1 \over 2\pi}\int_{-\infty}^{\infty}d\alpha~
\mbox{Tr}_R~e^{i\alpha (E-\hat{H})}=
{1 \over 2\pi}\int_{-\infty}^{\infty}d\alpha~
e^{i\alpha E}Z_R(i\alpha)~~~. 
\end{equation} 
The integration in (\ref{10.2}) can be performed \cite{Bateman}. For
$E\geq \mu_R$ one finds 
\begin{equation}\label{6.30}
\nu_R(E)=\delta(E-\mu_R)+\tilde{\nu}_R(E)~~~,
\end{equation}
\begin{equation}\label{6.30a}
\tilde{\nu}_R(E)=\left({\lambda_R \over E-\mu_R}\right)^{1/2}
I_1\left(2\sqrt{\lambda_R(E-\mu_R)}\right)~~~,
\end{equation}
where $I_1(x)$ is the modified Bessel function. The function
$Z_R(\beta)$ is obtained from $\nu_R(E)$ with the help of the Laplace 
transformation
\begin{equation}\label{6.29}
Z_R(\beta)=e^{-\beta\mu_R}+
\int_{\mu_R}^{\infty}e^{-\beta E}
\tilde{\nu}_R(E)~ dE~~~.
\end{equation}
It is possible to show that the spectrum of $\hat{H}_R$ is positive
$(\mu_R>0)$.

The function $w_R(E,\beta)$ describing the probability density  to find
the energy of the corresponding  canonical ensemble at the temperature
$\beta^{-1}$ in the energy interval between $E$ and $E+dE$ is
\begin{equation}\label{10.4}
w_R(E,\beta)=Z_R(\beta)^{-1}\nu_R(E)e^{-\beta E}\, .
\end{equation}
According to (\ref{10.2}), it is normalized to unity.

We shall be interested in the probability density at the Hawking
temperature $w_R(E)\equiv w_R(E,\beta_H)$. Since  $E_R-\mu_R\gg
\lambda_R^{-1}>0$ and we are interested in the energy region near
$E_R$, we can use the asymptotics of the Bessel functions to obtain
from (\ref{6.30a})
\begin{equation}\label{6.31}
w_R(E)\simeq {1 \over (4\pi)^{1/2}Z_R(\beta_H)} 
\left({\lambda_R \over (E-\mu_R)^3}\right)^{1/4}
e^{f_R(E)-\beta_H E}~~~,
\end{equation}
\begin{equation}\label{6.31a}
f_R(E)=2\sqrt{\lambda_R(E-\mu_R)}~~~.
\end{equation}
Then it is easy to show that function (\ref{6.31}) can be approximated
by a Gauss distribution with the center at $E=E_R$. One can check  with
the help of Eq.(\ref{6.91}) that near the maximum
\begin{equation}\label{10.5}
f_R(E)-\beta_H E\simeq\ln Z_R(\beta_H)- {(E-E_R)^2
\over \sigma^2_R}~~~,
\end{equation}
where $\sigma^2_R=4\lambda_R/\beta_H^3=g/24$.  So that we have
\begin{equation}\label{10.6}
w_R(E)\simeq {1 \over {\sigma_R\sqrt{\pi} }}
e^{-{(E-E_R)^2 \over \sigma^2_R}}~~~.
\end{equation}
The parameter $\sigma_R$ gives the width of the peak and it is of the
order of  magnitude of the mass-parameter, characterizing the cut-off
scale. It depends on the regularization scheme. For instance, in the
Pauli-Villars regularization \cite{DLM} $\sigma_R$ is the largest mass
of the auxiliary  Pauli-Villars fields. The appearance of the
ultraviolet cut-off in distribution (\ref{6.31}) is explained by the
fact that the function $Z_R(\beta)$  has the ultraviolet divergencies
which have to be  regularized. 

One can easily obtain the Rindler density of states $\nu_R(E)$ at the
peak of the probability distribution. Substituting (\ref{6.31}) into
(\ref{10.4}) one has
\begin{equation}\label{10.6a}
\nu_R(E_R)\sim \exp f_R(E_R)
=\exp S_R~~~,
\end{equation}
where $S_R={\pi\over 3}g {\cal A}^H$ and ${\cal A}^H=16\pi M^2$ is the
surface area of the black hole.  As expected the logarithm of
$\nu_R(E_R)$ is the entropy of the canonical ensemble of the Rindler
particles.

\section{Probability distribution and degeneracy of black hole states}
\setcounter{equation}0

Let us establish now the relation between  degrees  of freedom of
Rindler particles and excitation states of a black hole.  First of all
we note that by  using Eqs. (\ref{1.1}) and  (\ref{2.10}) one can
relate  the Rindler energy and entropy to  the total average energy
$\bar{E}$ and  Bekenstein-Hawking entropy $S^{BH}$   
\begin{equation}\label{6.7}
S_{R}=S^{BH}+\bar{Q}~~~,
\end{equation}
\begin{equation}\label{6.8}
E_R=\bar{E} +\beta_H^{-1}\bar{Q}~~~.
\end{equation}
Here
\begin{equation}\label{6.8b}
\bar{Q}=\langle \hat{Q}\rangle
=2\pi \sum_s \xi_s g(m^2_s){\cal A}^H
~~~.
\end{equation}
Since in the induced gravity  on Ricci-flat backgrounds $\bar{E}=0$ one
has
\begin{equation}\label{6.8a}
E_R=\beta_H^{-1}\bar{Q}~~~.
\end{equation}
Substituting this relation into  (\ref{6.93}) one  defines the
parameter $\mu_R$ in  the Rindler partition function (\ref{6.91}).

As we mentioned, at least some of the non-minimal  coupling constants
$\xi_s$ must be positive. In what follows we assume for simplicity that
all $\xi_s>0$, and hence the charge $\bar{Q}$ is positive. The
generalization to the case where this assumption is not satisfied is
straightforward.

Relations (\ref{6.7}) and (\ref{6.8}) indicate that only part of the
total energy and entropy of Rindler modes is responsible for thermal
characteristics of a black hole.  It suggests that the total system
described by Rindler modes consists in fact of two independent parts,
one is connected with "physical" degrees of freedom of the black hole,
and  the other is a subsystem of soft modes.  The soft modes do not
contribute to the  canonical energy.  In other words, one can add any
number of soft modes  to the given state without changing its Rindler
energy.  Therefore we identify the space of "physical" states  with the
space of Rindler states  modulo the  subspace of soft modes. We
demonstrate that this identification correctly reproduces the
degeneracy of the black hole mass.

In accordance with relations (\ref{6.7}) and (\ref{6.8}) and our
assumption the partition function obeys the factorization property  
\begin{equation}\label{6.9}
Z_R(\beta)=Z(\beta)~Z_{Q}(\beta)~~~.
\end{equation}
Here  $Z(\beta)$ and $Z_{Q}(\beta)$ are  partition functions for
"physical" and soft modes, respectively. Equations (\ref{6.7}) and
(\ref{6.8}) follow from (\ref{6.9}) provided
\begin{equation}\label{aaa}
\beta^{-1}_H\bar{Q}=-{\partial \over \partial \beta}
\ln Z_Q(\beta)|_{\beta=\beta_H}~~~,
\end{equation}
\begin{equation}\label{aab}
\bar{Q}=-\left(\beta
{\partial \over \partial \beta}-1\right)
\ln Z_Q(\beta)|_{\beta=\beta_H}~~~.
\end{equation}
We recall that $\bar{Q}$ and $Z_R$ are divergent and  so some
regularization in (\ref{6.9}) is supposed.  It is important  that all
three partition functions  that enter this relation  are regularized by
using the same regularization scheme.  Eqs. (\ref{aaa}) and (\ref{aab})
imply that
\begin{equation}\label{aac}
Z_Q(\beta_H)=1~~~.
\end{equation}

One can introduce the density of states $\nu(E)$ for $Z(\beta)$ as 
\begin{equation}\label{a10}
\nu(E)=
{1 \over 2\pi}\int_{-\infty}^{\infty}d\alpha~
e^{i\alpha E}Z(i\alpha)~~~, 
\end{equation} 
so that
\begin{equation}\label{a11}
Z(\beta)=\int_{\mu}^{\infty}dE e^{-\beta E}\nu(E)~~~.
\end{equation} 
Similar relations are used to define $\nu_Q(E)$ in terms of
$Z_Q(\beta)$. We assume that $\nu(E)$ and $\nu_Q(E)$ are non-vanishing
only for $E\ge \mu$ and $E\ge\mu_Q$, respectively, where $\mu$ and
$\mu_Q$ are some constants. Their value will be specified later. The
factorization property (\ref{6.9}) implies the following relation
between the densities of states
\begin{equation}\label{a12}
\nu_R(E)=\int_{\mu}^{E-\mu_Q}\nu(E')
\nu_Q(E-E') dE'~~~.
\end{equation}
We can also define by the relations similar to (\ref{10.4})  the
probability distributions   $w(E)$ and $w_Q(E)$ for each of the
subsystems at $\beta=\beta_H$ .  Then equation (\ref{a12}) and 
factorization formula (\ref{6.9}) result in the following relation
\begin{equation}\label{a13}
w_R(E)=\int_{\mu}^{E-\mu_Q}dE' w(E') w_Q(E-E') ~~~.
\end{equation}

Important properties of the distribution  $w_Q$   for the soft modes
are determined by Eqs.  (\ref{aaa}) and (\ref{aab}). Namely let us
write 
\begin{equation}\label{a13a}
\nu_Q(E)=\exp f_Q(E)~~~,
\end{equation}
and assume that the function $f_Q(E)$ grows  at infinity slower  than
$E$. In this case the probability distribution  $w_Q(E)$ has a maximum
at the point $E_Q$ where $f_Q'(E_Q)=\beta_H$. Near this maximum
\begin{equation}\label{a14}
w_Q(E)\sim {1\over \sigma_Q\sqrt{\pi}}\exp\left[-{(E-E_Q)^2\over 
\sigma_Q^2}\right]~~~,
\end{equation}
where $\sigma_Q^{-2}={1\over 2}|f_Q''(E_Q)|$. Note that Eq.(\ref{aaa})
can be rewritten as
\begin{equation}\label{a15}
\beta_H^{-1}\bar{Q}=\int_{\mu_Q}^{\infty}dE ~E~ w_Q(E)~~~.
\end{equation}
So by using the Gaussian approximation (\ref{a14}) we get 
\begin{equation}\label{a16}
E_Q\simeq \beta_H^{-1}\bar{Q}=E_R~~~      
\end{equation}
and $E_Q$ turns out to be the average energy for  the canonical
ensemble of the soft modes. On the other hand, Eq.(\ref{aac}) gives 
\begin{equation}\label{a16a}
f_Q(E_Q) \simeq \beta_H E_Q~~~.
\end{equation}
Consequently the density  number of states $\nu_Q(E)$ at the peak
$E=E_Q$ is
\begin{equation}\label{a16b}
\nu_Q(E)\simeq \exp (\beta_H E_Q)\simeq
\exp\bar{Q}~~~.
\end{equation}
According to the last equation the Noether charge $\bar{Q}$ can be
interpreted as the entropy of the soft modes.

In the Gaussian  approximation (\ref{a14}) equation (\ref{a13}) takes
the form
\begin{equation}\label{a17}
{1\over \sigma_R\sqrt{\pi}}~e^{-{(E-E_R)^2\over \sigma_R^2}}=
\int dE'~ w(E')~ {1\over \sigma_Q\sqrt{\pi}}~e^{-{(E-E'-E_Q)^2\over 
\sigma_Q^2}}~~~,
\end{equation}
where $E_R\simeq E_Q$. It follows from (\ref{a17}) that the
probability  distribution $w(E)$  also has  the Gaussian form
\begin{equation}\label{a18}
w(E)\sim {1\over \sigma\sqrt{\pi}}e^{-{E^2\over \sigma^2}}~~~,
\end{equation}
with the dispersion $\sigma$
\begin{equation}\label{a19}
\sigma^2=\sigma_R^2-\sigma_Q^2~~~.
\end{equation}
Now by taking into account Eq.(\ref{aac}) and the fact  that the
distribution of the soft modes  has the peak at $E=E_Q$ we find from
(\ref{a12})
\begin{equation}\label{densities}
\nu_R(E)\sim 
\nu(E-E_Q)\nu_Q(E_Q)~~~.
\end{equation}
The distribution (\ref{a18}) of  the "physical" modes is centered at
$E=0$.  The density of states at the maximum of the function  $w(E)$,
according to (\ref{densities}), is
\begin{equation}\label{a20}
\nu(0)\sim{\nu_R(E_R)\over \nu_Q(E_Q)}\sim \exp (S_R-\bar{Q})~~~.
\end{equation}
By using  relation (\ref{6.7})  one can rewrite this expression as
\begin{equation}\label{a21}
\nu(0)\sim \exp {S^{BH}}=\exp {{\cal A}^H\over 4G}~~~.
\end{equation}
Thus we proved that the spectrum of the "physical" states correctly
reproduces the degeneracy of black hole mass levels.  Therefore the
identification of  the black hole states with the "physical" states is
justified. As the result the distribution of the black  hole mass
induced by quantum fluctuations of the  constituents is centered near
the average value $M$ and has the width $\sigma$.

The following arguments can be used now to obtain an additional
information concerning the width $\sigma$.  By  assuming that the
function $Z_Q(\beta)$ has the temperature asymptotic similar to that of
$Z_R(\beta)$, see Eq.(\ref{6.91}), one can write
\begin{equation}\label{a22}
Z_Q(\beta)=\exp(-\beta \mu_Q+\beta^{-1}\lambda_Q)~~~.
\end{equation}
The parameters $\mu_Q$ and $\lambda_Q$ can be found from
Eqs.(\ref{aaa}) and (\ref{aab})
\begin{equation}\label{a23}
\lambda_Q={1\over 2}\beta_H \bar{Q}\, ,\hspace{0.5cm}
\mu_Q={1\over 2\beta_H }\bar{Q}~~~.
\end{equation}
Using these relations we get
\begin{equation}\label{a23a}
Z_Q(\beta)=\exp\left[-{1\over 2}\left({\beta\over \beta_H}
-{\beta_H\over \beta}\right)\bar{Q}\right] ~~~.
\end{equation}
For this partition function $\sigma_Q^2=4\lambda_Q\beta_H^{-3}=\frac 14
\sum_s\xi_s g_s$ . Hence, according to (\ref{a19}) and (\ref{yyy}),
\begin{equation}\label{a24}
\sigma^2 ={1 \over 24} \sum_i g(m_i^2)-\frac 14 \sum_s \xi_s g(m^2_s)
={1 \over 32\pi G}~~~.
\end{equation}
In other words, the width $\sigma$     of the probability distribution
of the black hole  states does not depend on the regularization
ambiguity,  it is finite and proportional  to the Planck mass $m_{Pl}=
G^{-1/2}$.

The factorization property (\ref{6.9}) together with (\ref{a23a})
implies that $Z(\beta)$ has the same form as $Z_R(\beta)$ and
$Z_Q(\beta)$ and can be written explicitly as
\begin{equation}\label{a25}
Z(\beta)=\exp\left[\frac 12\left({\beta \over \beta_H}
+{\beta_H \over \beta}\right) S^{BH}\right]~~~.
\end{equation}
The important property of the partition function of the "physical"
degrees of freedom in the induced gravity is that  it is defined
entirely by the ultraviolet finite quantity $S^{BH}$ (at least in the
one-loop approximation).  So the function $Z(\beta)$, even if it is
taken off-shell, i.e. for an arbitrary temperature $\beta^{-1}$,   is
well defined and ultraviolet finite.

\section{Soft modes and fluctons} \setcounter{equation}0

The degrees of freedom which enable one to single out "physical" states
from the space of Rindler states are associated with the soft modes.
Yet an  explicit formulation of the black hole statistical-mechanics in
terms of "physical" degrees of freedom is a non-trivial problem. To
some extend  this reminds the  situation in the gauge theories where in
general the constraints cannot be resolved explicitly. In many cases,
however, it is sufficient to describe the physical space indirectly as
a factorization of an extended space over the group of gauge
transformations. In the functional integral  such a factorization is
realized by introducing the  Faddeev-Popov ghosts.

Let us make some additional remarks  concerning the subsystem of the
soft modes. We saw that these modes are located in the nearest vicinity
of the horizon, thus their physics is two dimensional by its nature.
Moreover we will demonstrate now that the charge $\bar{Q}$ can be
expressed as an effective action of a two-dimensional quantum theory. 

To this aim we use representation (\ref{3.18}) for the correlator of
the scalar field  on the bifurcation surface $\Sigma$. In the
Hartle-Hawking state ($\beta=\beta_H$)  formula (\ref{3.18}) reads
\begin{equation}\label{3.19}
{\cal G}(z,z')=
-{1 \over 4\pi}\langle z| 
\ln((-\nabla_\Sigma^2+m^2)\epsilon^2)|z'\rangle~~~,
\end{equation}
where $-\nabla_\Sigma^2$ is the Laplace operator on $\Sigma$. Thus one
can write
\begin{equation}\label{fluc1}
2\pi\int_{\Sigma}\langle\hat{\phi}(z)^2
\rangle~d^2 z = 
2\pi\int_{\Sigma}{\cal G}(z,z)~d^2 z=
-\frac 12\ln\det((-\nabla_\Sigma^2+m^2)\epsilon^2)~~~.
\end{equation}
The quantity  $W_\chi=\frac 12\ln\det[(-\nabla_\Sigma^2+m^2)
\epsilon^2]$  is identical to the effective action for a
two-dimensional quantum field $\chi$ defined on $\Sigma$.  The
functional $W_\chi$ can be rewritten as the Euclidean functional
integral over the field $\chi$
\begin{equation}\label{fluc2}
e^{-W_\chi}=
\int D[\chi]
\exp\left[-\frac 12\int_{\Sigma}
((\nabla_\Sigma\chi)^2+m^2\chi^2)~d^2 z\right]
~~~.
\end{equation}
We call $\chi$ {\it flucton} field to distinguish it from the original
scalar field $\phi$. The flucton field is the free field with the same
mass  $m$ as the 4D field $\phi$.  According to  Eq.(\ref{3.19}), the
quantum theory of fluctons is completely defined by the 4D correlator
${\cal G}(z,z')$  on $\Sigma$. 

From (\ref{fluc2}) one obtains the representation for the charge
$\bar{Q}$
\begin{equation}\label{fluc3}
\bar{Q}=2\pi\xi\int_{\Sigma}\langle\hat{\phi}(z)^2
\rangle~d^2z=-\xi W_\chi ~~~.
\end{equation}
As was shown in Section 7, the quantity  $\bar{Q}$ coincides with the
entropy of the ensemble of the soft modes (see Eq. (\ref{a16b})). On
the other hand, the integral (\ref{fluc2}) can be interpreted as
microcanonical partition function of the flucton fields and  $W_\chi$
as the microcanonical free energy.  Consequently, $W_\chi=-S_{\chi}~$,
where $S_{\chi}$  is the entropy of fluctons.  These arguments enable
one to represent the Noether charge  in a pure statistical mechanical
form
\begin{equation}\label{fluc4}
\bar{Q}=-\xi W_\chi=\xi S_{\chi}~~~
\end{equation}
and relate it to the entropy of a  microcanonical ensemble of
two-dimensional fields $\chi$ on $\Sigma$. The constant of the
non-minimal coupling $\xi$ plays in (\ref{fluc4}) a role of the
effective number of the flucton fields.

To derive Eq. (\ref{fluc4}) we used the Rindler approximation for the
black hole geometry. It is possible to show (see Appendix) how to
extend this two-dimensional interpretation to the case of arbitrary
black hole backgrounds.

\section{Discussion} \setcounter{equation}0

We  make now some general remarks concerning the derivation of the
Bekenstein-Hawking entropy by counting the degrees of freedom of
constituents in the induced gravity. First, let us compare the induced
gravity with other theories. In general case one has for the observable
value of the Newtonian constant $G$ the following expression
$G^{-1}=G^{-1}_{bare}+G^{-1}_{q}$, where $G^{-1}_{q}$  is a (one-loop)
quantum correction to the initial bare constant $G_{bare}^{-1}$. In
order to obtain the finite value of $G$ one usually begins with the
infinite quantity $G^{-1}_{bare}$   which absorbs the ultraviolet
divergences. As the result in the expression for the black hole entropy
besides the part ${\cal A}^H/(4G_{q})$ that can be connected with
statistical mechanics there is always  the term ${\cal
A}^H/(4G_{bare})$ having no clear statistical-mechanical meaning. In
the induced gravity $G^{-1}_{bare}=0$ and  this problem is solved
automatically.

In order to have the correct Einstein low energy  gravity with the
finite observable Newton constant one must impose special constrains on
the parameters of the fields inducing the gravity. In our particular
model these requirements are satisfied because of the presence of the
non-minimally coupled fields. The same set of constraints guarantees
that the induced entropy of the black hole is also finite and 
coincides with the Bekenstein-Hawking entropy $S^{BH}$. The entropy 
$S^{BH}$ can be obtained from the statistical-mechanical entropy of the
constituents by subtracting the  Noether charge $Q$ of the
non-minimally coupled fields.  The same quantity $Q$ determines the 
difference between the energy  $E$ of the fields in the black hole
exterior and the value of their Hamiltonian (canonical energy) $H$. We
showed that there exist a set of states (soft modes) that  contribute
to $Q$ but do not contribute to $H$,  so that the Hamiltonian for the
Rindler particles  is degenerate. By using the  factorization of the
space of states of the Rindler particles with respect to the subspace
of soft modes  we obtained the degeneracy of black hole states
responsible for the Bekenstein-Hawking entropy.

This mechanism is universal in the sense that it does not depend on the
concrete choice of the set of scalar and spinor constituents and their
properties provided the general constraints are satisfied. The concrete
model of the induced gravity may differ from the one considered in this
paper, and  may contain, for example, finite or infinite number of
fields of higher spins. However our consideration indicates that it is
quite plausible  that the same mechanism still works.

The  obtained statistical-mechanical representation of black hole
entropy in the induced gravity does not depend on the particular
structure of the theory at the Planck energies. All the information
about field species and masses of the heavy constituents in the
low-energy limit is compressed in the Newton constant (\ref{i10}). So
any two microscopically different theories, that induce at low energies
the same theory of gravity, predict the same entropy $S^{BH}$ for a
black hole. The details and the way of statistical-mechanical
calculations of $S^{BH}$ may depend on the type of the theory, but the
results of calculations will coincide. The assumption that this happens
in all theories having the Einstein gravity in the low energy limit was
called in Ref.\cite{FFZ3} the {\it low-energy censorship conjecture}.

It should be emphasized that we do not consider   Sakharov's approach
as a version of the final theory of quantum gravity. Certainly, it
cannot compete with the superstring theory, which is considered as a
modern candidate for quantum gravity theory. There are many indications
that the superstring  models give the correct answer for the black hole
entropy by counting string degrees of freedom. But we would like to
stress that the string calculations essentially use supersymmetry and
usually deal with the black holes close to the extreme ones. Moreover 
for each model and type of a black hole the proof of the corresponding
result requires new calculations and is often considered as a miracle
(see e.g \cite{Horowitz}).

Since the thermodynamical characteristics of macroscopical  black holes
are determined by  a low-energy effective theory of gravity it is
reasonable to suggest that there exists some mechanism that guarantees
this universality. The models of induced gravity might be interesting
as some kind of the phenomenological models in which many details
concerning the underlying  microscopical theory are lost, and only a
few of  its most important features are preserved.  In particular in
Sakharov's approach, as well as in the string theory, the gravity is
the induced phenomenon and the  Newton constant is ultraviolet finite. 
Our analysis indicates that namely these two features  are sufficient
for  the statistical-mechanical explanation of the Bekenstein-Hawking
entropy. That is why we believe that the proposed mechanism of black
hole entropy generation may be of more general interest.

\vspace{12pt} {\bf Acknowledgements}:\ \ This work was supported  by
the Natural Sciences and Engineering Research Council of Canada.

\newpage \appendix
\section{Noether charge of a non-minimally coupled scalar field as a 2D
effective action} \setcounter{equation}0

Let us consider the correlator of quantum scalar field $\phi$, see
(\ref{2.3}),    on the bifurcation surface $\Sigma$  of the black hole
horizons
\begin{equation}\label{4.1}
\langle\hat{\phi}(x(z))\hat{\phi}(x(z'))
\rangle={\cal G}(z,z')~~~.
\end{equation}
As earlier ${\cal G}(z,z')$ is the value of the Green $G(x,x')$
function with its both arguments taken on $\Sigma$. The arguments $z$
and $z'$ are the coordinates   of the points $x$ and $x'$ on $\Sigma$,
analogous to the coordinates $z$ on the Rindler horizon  in  metric
(\ref{3.1}). Here we will be interested in function (\ref{4.1}) on a
general black hole  background. We assume that the quantum state is the
Hartle-Hawking vacuum. Then function ${\cal G}(z,z')$ has an Hadamard
form and  one can write for it  the following Schwinger-DeWitt
representation \cite{DeWitt} 
\begin{equation}\label{4.2}
{\cal G}(z,z')=\triangle^{1/2}(z,z') \int_{\delta}^{\infty}ds 
{1 \over (4\pi s)
^2}~e^{-{\sigma^2(z-z') \over 4s}-m^2s}
\left(1+a_1(z,z')s+a_2(z,z')s^2+...\right)~~~.
\end{equation}
Here $\delta$ is ultraviolet cut-off parameter,  $\sigma(z-z')$ is the
4D geodesic distance between the points, and 
$$
\triangle(x,x')=-[g(x)g(x')]^{-1/2}
\det\left(\frac 12{ \partial^2 \sigma(x,x') \over 
\partial x^{\mu}\partial x'^{\nu}}\right)
$$
is the Van Vleck determinant. $a_i(x,x')$ are the coefficients of the
asymptotic expansion of the heat kernel of the scalar wave operator 
$L=-{\,\lower0.9pt\vbox{\hrule \hbox{\vrule height 
0.2 cm \hskip 0.2 cm \vrule height 0.2 cm}\hrule}\,}+m^2+\xi R$,
see Eq. (\ref{2.3}). The integration contour in  (\ref{4.2}) can be
chosen real.

It is important to note that a  two-dimensional geodesic on $\Sigma$ is
also a geodesic in the enveloping space-time. Indeed, because $\Sigma$
is the fixed set of the Killing field, both second fundamental forms 
of $\Sigma$ vanish.  This is a necessary and sufficient condition for
$\Sigma$ to be a totally geodesic surface \cite{Eisenhart}.  Thus one
can substitute instead of  $\sigma(z,z')$ in (\ref{4.2}) the 2D
geodesic distance $\sigma_{\Sigma}(z,z')$ on $\Sigma$. Let us consider
now a two-dimensional operator 
\begin{equation}\label{4.3}
O_{\Sigma}=-\nabla^2_{\Sigma}+m^2+
V[{\cal R}]~~~,
\end{equation}
where $-\nabla^2_{\Sigma}$ is the Laplacian on $\Sigma$, and $V[{\cal
R}]$ is a "potential" which may depend on the scalar curvature
$R_\Sigma$ of $\Sigma$, and an external geometry in the vicinity of
this surface. The matrix element of the heat kernel of $O_\Sigma$ has
the following asymptotic form 
\begin{equation}\label{4.4}
\langle z|e^{-s O_{\Sigma}}|z'\rangle\simeq
{\triangle^{1/2}_\Sigma(z,z') \over (4\pi s)}~
e^{-{\sigma_{\Sigma}^2(z,z') \over 4s}-m^2s}
\left(1+a_{\Sigma,1}(z,z')s+a_{\Sigma,2}(z,z')s^2
+...\right)~~~,
\end{equation}
where $a_{\Sigma,i}(z,z')$ and  $\triangle^{1/2}_ \Sigma(z,z')$ are the
heat kernel coefficients and the Van Vleck determinant on $\Sigma$,
respectively.

Suppose now that the potential $V[{\cal R}]$ of the operator
(\ref{4.3}) on $\Sigma$ can be chosen  so that
\begin{equation}\label{4.4b}
a_{i}(z,z')\simeq a_{\Sigma,i}(z,z')~~~.
\end{equation}
Then by comparing (\ref{4.4}) with  (\ref{4.2}) and taking into account
that $\Sigma$ is a totally geodesic surface one finds that
\begin{equation}\label{4.5}
{\cal G}(z,z')\simeq{1 \over 4\pi}e^{f(z,z')}
\int_{\delta}^{\infty}
{ds \over s}
\langle z|e^{-s O_{\Sigma}}|z'\rangle
=-
{1 \over 4\pi}~e^{f(z,z')}
\langle z|\ln O_{\Sigma}|z'\rangle~~~.
\end{equation}
Here the logarithm of the operator is understood as a regularized
quantity, in the same way as the correlator (\ref{4.2}). The function
$f(z,z')$ is defined as
\begin{equation}\label{4.5b}
e^{f(z,z')}={\triangle^{1/2}(z,z') \over 
\triangle^{1/2}_\Sigma(z,z')}~~~.
\end{equation}
For a spherical horizon the curvature decomposition of $f(z,z')$ looks
as 
\begin{equation}\label{4.5c}
f(z,z')={\sigma_{\Sigma}^2(z,z') 
\over 24}(R-R_{\mu\nu}n^{\mu}_in^{\nu}_i-R_{\Sigma})+...~~~,
\end{equation}
where $n^\mu_i$ are two unit orthogonal vectors  which are normal to
$\Sigma$.   The r.h.s. of (\ref{4.5})  enables one an interpretation in
terms of two-dimensional quantum theory of flucton field $\chi$ on
$\Sigma$. Indeed, $f(z,z)=0$ and
\begin{equation}\label{4.7}
\frac 12 \int_{\Sigma}\langle z|\ln O_{\Sigma}|z \rangle 
~\sqrt{\gamma}~d^{2}z=\frac 12 \ln \det(
-\nabla^2_{\Sigma}+m^2+V[{\cal R}] )
\equiv W_\chi[\gamma]~~~.
\end{equation}
The functional $W_\chi[\gamma]$ has the meaning of an   effective
action. It is expressed in terms the  Euclidean path integral as
\begin{equation}\label{a4.10}
e^{-W_\chi[\gamma]}=\int D[\chi]
\exp\left[-\frac 12\int_{\Sigma}(\chi^{,i}\chi_{,i}
+(m_s^2+V[{\cal R}])\chi^2)
\sqrt{\gamma}~d^2 z\right]
~~~.
\end{equation}
where $D[\chi]$ is a covariant measure. From (\ref{4.7}) one finds the
representation of the Noether charge in terms of  the effective action
of flucton field $\chi$
\begin{equation}\label{a4.9a}
\bar{Q}=-\xi W_\chi[\gamma]~~~.
\end{equation}
Formula (\ref{a4.9a}) is the generalization of  Eq. (\ref{fluc3})
obtained in Section 8   in the Rindler approximation.

Let us note that in the case of the Rindler space $V[{\cal R}]=0$,
$f(z,z')=0$ and basic equality (\ref{4.5})   holds exactly.   However,
developing flucton theory on a general  background is a more  difficult
problem. Such a theory may even not exist in a local form, if the
relations (\ref{4.4b}) are not satisfied for $2D$  dimensional elliptic
operators. However, this is not an obstruction for the models of  the
induced gravity   based on the assumption that the Compton wave length
$\lambda$ of the heavy constituents is much smaller than the curvature
radius of the background space. Because the field correlators vanish
when  $\sigma(z,z')\gg \lambda$, it is possible  to satisfy Eq.
(\ref{4.4b}) only approximately for the first coefficients and in some
order in curvature. Moreover, to find the contribution to the
Bekenstein-Hawking entropy (\ref{1.1}) it is  sufficient to calculate
$\langle\hat{\phi}^2 \rangle$ by neglecting the curvature effects at
all.  That is why the Rindler approximation was justified in our
analysis.


\newpage

\begin{thebibliography}{000}
\bibitem{Beke:72} J.D. Bekenstein, Nuov. Cim. Lett. {\bf 4}   (1972)
737.
\bibitem{Beke:7374} J.D. Bekenstein, Phys. Rev. {\bf D7} (1973) 2333; 
Phys. Rev. {\bf D9}  (1974) 3292.
\bibitem{Hawk:75} S.W. Hawking, Comm. Math. Phys. {\bf 43} (1975) 199.
\bibitem{Hooft:85} G.'t Hooft, Nucl. Phys. {\bf B256} (1985) 727.
\bibitem{Sorkin} L. Bombelli, R. Koul, J. Lee, and R. Sorkin, Phys.
Rev. {\bf D34} (1986) 373.
\bibitem{Srednicki} M. Srednicki, Phys. Rev. Lett. {\bf 71} (1993) 666.
\bibitem{FroNo} V. Frolov and I. Novikov, Phys. Rev. {\bf D48} (1993)
4545.
\bibitem{Frol:95} V.P. Frolov, Phys. Rev. Lett. {\bf 74} (1995) 3319.
\bibitem{SU} L. Susskind and J. Uglum, Phys. Rev. {\bf D50} (1994) 2700.
\bibitem{CW} C. Callan and F. Wilczek, Phys. Lett. {\bf B333} (1995) 55.
\bibitem{BCZ} A.A. Bytsenko, G. Cognola, and S. Zerbini, Nucl. Phys.
{\bf B458} (1996) 267.
\bibitem{BZ} A. Zelnikov  and  A. Barvinskii, {\it Density Matrix of a
Black Hole and Bekenstein-Hawking Entropy}, to appear.
\bibitem{Jacobson} T. Jacobson, {\it Black Hole Entropy and Induced
Gravity}, preprint gr-qc/9404039.
\bibitem{Sakh} A.D. Sakharov, Sov. Phys. Doklady, {\bf 12} (1968) 1040,
Theor. Math. Phys. {\bf 23} (1976) 435.
\bibitem{Adler} S.L. Adler, Rev. Mod. Phys. {\bf 54} (1982) 729.
\bibitem{FFZ3} V.P. Frolov, D.V. Fursaev, and A.I. Zelnikov, Nucl.
Phys. {\bf B486} (1997) 339.
\bibitem{Wald:93} R.M. Wald, Phys. Rev. {\bf D48} (1993) R3427.
\bibitem{JKM} T.A. Jacobson, G. Kang, and R.C. Myers, Phys. Rev. {\bf
D49} (1994) 6587.
\bibitem{Mayo} A.E. Mayo and J.D. Bekenstein, Phys. Rev. {\bf D54}
(1996) 5059.
\bibitem{Solod:95} S.N. Solodukhin, Phys. Rev. {\bf D52} (1995) 7046.
\bibitem{DLM} J.-G. Demers, R. Lafrance, and R.C. Myers, Phys. Rev.
{\bf D52}  (1995) 2245.
\bibitem{LW} F. Larsen and F. Wilczek, Nucl. Phys. {\bf B458} (1996)
249.
\bibitem{Hotta} M. Hotta, T. Kato and K. Nagata, {\it A Comment on
Geometric Entropy and Conical Space}, gr-qc/9611058, V. Moretti, {\it
Geometric Entropy and  Curvature Coupling in Conical Spaces: $\zeta$
Function Approach}, hep-th/9701099. \bibitem{Solod:9612} S.N.
Solodukhin, {\it Non-Minimal Coupling and Quantum Entropy of a Black 
Hole}, hep-th/9612061.
\bibitem{MTW} C.W. Misner, K.S. Thorne, and J.A. Wheeler, {\it
Gravitation}, W.H. Freeman and Company, New York.
\bibitem{Belgiorno:9612} F. Belgiorno and S. Liberati, {\it Black Hole
Thermodynamics, Casimir Effect and Induced Gravity}, gr-qc/9612024.
\bibitem{Nelson} W. Nelson, Phys. Rev. {\bf D50}  (1994) 7400.
\bibitem{FS:95} D.V. Fursaev and S.N. Solodukhin, Phys. Rev. {\bf D52}
(1995) 2143.
\bibitem{WaldIyer:95} V. Iyer and R.M. Wald, Phys. Rev. {\bf D52}
(1995) 4430.
\bibitem{Takagi} S. Takagi, Progress of Theoretical Physics Supplement
{\bf 88} (1986).
\bibitem{Hawk:76} S.W.Hawking, Phys. Rev. {\bf D13} (1976) 191.
\bibitem{York:86} J.W. York, Phys. Rev. {\bf D33} (1986) 2092.
\bibitem{BCH} J.M. Bardeen, B. Carter and S.W. Hawking, Commun. Math.
Phys. {\bf 31} (1973) 161.
\bibitem{Bateman} H. Bateman and A. Erdelyi, {\it Tables of Integral
Transformations}, v.1, New York, McGraw-Hill Book Company, Inc., 1954.
\bibitem{DeWitt} B.S. DeWitt, {\it Dynamical Theory of Groups and
Fields}, Gordon and Breach, New York 1965.
\bibitem{Eisenhart} L.P. Eisenhart, {\it Riemannian  Geometry},
Princeton University Press, 1966, Princeton. 
\bibitem{Horowitz} G.T. Horowitz, {\it The Origin of Black Hole Entropy
in String Theory}, gr-qc/9604051.

\end{thebibliography}
 \end{document}